# Magnetic properties of self-organized lateral arrays of (Fe,Ag)/Mo(110) nanostripes

B. Borca, O. Fruchart and C. Meyer*

Laboratoire Louis Néel, (CNRS, Université Joseph Fourier et Institut National Polytechnique de Grenoble), 25 Avenue des Martyrs, 38042 Grenoble, France

**Abstract**

We report the fabrication of self-organized arrays of Fe nanostripes with a period of 3.5 nm, by sequential deposition of Fe and Ag on Mo(110). The wires display a strong in-plane uniaxial anisotropy along their length, and are superparamagnetic above $T_B=185\pm15$K. The large value of nucleation volumes, inferred from the analysis of the thermal dependence of coercivity below $T_B$, suggests the existence of interactions between the wires.

*Corresponding author email: meyer@grenoble.cnrs.fr

**I. Introduction**

Artificially structured magnetic materials are of growing importance for fundamental and applied research in magnetism. Self-organization on surfaces is a powerful approach to fabricate nanostructures with properties tailored at the atomic scale[1]. This study focuses on lateral arrays of alternate nanowires of Ag and Fe that arise spontaneously upon deposition of these two elements on Mo(110). This system was first mentioned by Tober *et al.*[2,3] using molecular beam epitaxy (MBE). The formation of alternating Fe and Ag stripes parallel to Mo[001] with a few nm periodicity was explained with a modulated misfit strain relaxation along $[1\bar{1}0]$, positive for Fe and negative for Ag. This phenomenon was shown to exist up to four atomic layers (ALs). The system was found to be superparamagnetic at room temperature, and a weak in-plane CPP magnetoresistance $\Delta R/R=0.88\%$ was found at 2.7K. The purpose of our study is twofold: i) defining the growth parameters in the case of ultra-high vacuum pulsed laser deposition (UHV-PLD) to obtain these systems, ii) investigating the magnetic properties of the arrays as a function of temperature.

**II. Growth of the films**

The epitaxial films were grown in an ultra high vacuum deposition setup providing a base pressure of $5 \cdot 10^{-11}$ Torr. The evaporation from metallic targets is achieved by a Nd-YAG laser at 532 nm, with a pulse duration of 10 ns at a repetition rate of 10 Hz. The surface of the sample is monitored by Reflection High Energy Electron Diffraction (RHEED). The morphology of the surface is analyzed after deposition by scanning tunneling microscopy (STM). The nanostructures are grown on a Mo(110) 8 nm-thick buffer layer deposited on a $Al_2O_3(11\bar{2}0)$ sapphire substrate. The Mo surface is atomically flat with terraces larger than 200 nm, whose step orientation depends on the miscut of the substrate[4].

The (Fe,Ag)/Mo(110) nanostructures were grown at 420K by a series of $n$ sequential depositions of 0.5 AL Ag followed by 0.5 AL Fe ($n$ = 1 - 12). The deposition rate was 0.2Å/mn for Ag and 0.4Å/mn for Fe. In agreement with reference 2, we observe up to 4AL the formation of alternate stripes of Ag and Fe, both roughly aligned along Mo[001], with an average period of 3.5 nm along Mo$[1\bar{1}0]$. Beyond 4AL, a 2D nanostructuration takes place. Figure 1 summarizes this evolution as observed on the STM images taken on a wedge sample fabricated with a mask translated during the course of growth. This evolution can be understood by the limited vertical extension of the stress relaxation process. More details about the interpretation of the growth process will be published in a separate paper[5].

Figure 2 represents the STM image of an array of nanostripes with an equivalent thickness of 4AL (Fe,Ag). The 3.5nm average lateral periodicity of the corrugation is evidenced by the cross-section shown in the inset. These 3.5nm correspond to 15 to 16

Mo atomic rows along the $[1\bar{1}0]$ direction, which is equivalent to 8 Ag + 8 Fe atomic rows, recalling the stoichiometry is 1:1. Note that despite the good quality of the films, the surface is not atomically-flat on a large scale, implying a distribution of local coverage.

**III. Magnetization measurements on $(Ag,Fe)_4$**

In the present paper, we focus on the magnetic properties of the 4 AL sample. SQUID measurements were performed *ex situ* on Mo-capped films, with a maximum field of 5 Tesla and at variable temperature. Hysteresis was conducted along two in-plane directions: parallel to Fe[001] i.e. along the wires, and parallel to Fe$[1\bar{1}0]$ i.e. perpendicular to the wires. The data displayed have been processed to subtract the Sapphire diamagnetic susceptibility, which is of the same order of magnitude as the ferromagnetic signal. The value of susceptibility to subtract was estimated from the slope of the raw data towards high fields, where we expect the ferromagnetic sample to be saturated.

As can be seen on Figure 3a, the in-plane hard axis is $[1\bar{1}0]$, with a saturation field around 1T at 10K. The easy axis of magnetization lies along [001], along which the remanent magnetization Mr/Ms equals 0.77, with a coercive field $\mu_0 Hc$ =0.32 T. At room temperature, the uniaxial anisotropy is reduced (Figure 3b), while no coercivity is observed along the easy axis, suggesting a superparamagnetic behavior. We can exclude a paramagnetic behavior, because the initial susceptibility is high.

Let us first discuss the anisotropy. In the following we use the bulk iron magnetization value, $M_S(0K)=1750$ kA/m, as a first approach. Then, the total anisotropy energy density can be derived as the area above the hard axis curve, yielding: $E_{exp}=5.6\pm0.5\times10^5$ J/m$^3$ at 10K, or in terms of a mean uniaxial anisotropy field $\mu_0 H_A = 2K/M_s$, 0.65±0.05T. Let us consider a perfect array of parallel and infinitely-long Fe wires, 0.8nm-thick, 1.5nm-wide, with a 2nm gap along $[1\bar{1}0]$, to estimate average parameters for the real array inevitably displaying some distribution of size and orientation. The in-plane demagnetizing coefficient of an isolated wire is $N_{wire} \approx 0.36$, while that of the present array is reduced to $N_{array} \approx 0.31$ owing to dipolar coupling between the wires. Taking into account the distributions of the real samples, we thus estimate the dipolar energy along $[1\bar{1}0]$ to be $E_d \approx (6\pm1)\times10^5$ J/m$^3$, thus being very similar to $E_{exp}$. This means that all other sources of magnetic anisotropy cancel out. These sources are of microscopic origin, both interfacial and magnetoelastic. Let us consider a Fe continuous film for comparison. Mo/Fe (4AL ≈0.8nm)/Mo(110) films[6] also display a uniaxial anisotropy energy in-the-plane, surprisingly of the very same magnitude $6.0\times10^5$ J/m$^3$. This anisotropy was ascribed predominantly to a Néel-type interfacial anisotropy between Fe and Mo[6]. In our present case additional interfaces occur between Fe and Ag, with a geometry perpendicular-to-the-plane, and a bcc(110)\fcc(112) matching of atomic planes. Fe/Ag interfaces are thought to favor the alignment of magnetization along the interface[7], and thus should add up to the Mo/Fe/Mo contribution. As we evidence experimentally a nearly zero total microscopic energy, this means that additional contributions balance the above two mentioned. A first contribution is the ridges of the Fe wires, with a Ag,Mo environment. Such ridges, of one-dimensional nature, should be associated to a magnetic anisotropy energy of importance here, as the cross-over between 2D to 1D anisotropy lies around a few atoms in width for stripes[8]. A second contribution is the magnetoelastic energy. Unfortunately the strain distribution is complex: large and tensile along [001], modulated along [$1\bar{1}0$], and presumably compressed along [110]. Besides, the strain lies well beyond the validity of linear magnetoelasticity, so that no quantitative values can be predicted. The contribution between interfacial anisotropy of lateral interfaces, and magnetoelastic energy, can therefore not be disentangled.

Let us now discuss magnetization reversal. At 10K the ratio $H_C/H_A$ equals 0.5, not far from 1. This suggests that the micromagnetic features of nucleation events triggering magnetization reversal can be reasonably described by the coherent rotation model, of course with nucleation volumes $V_n$ much smaller than the total volume of the film[9]. Information about intrinsic properties of the sample should therefore be gained by analyzing the thermal dependence of coercivity. Figure 4 shows variable-temperature hysteresis loops, while the temperature-dependant coercivity $H_C(T)$ is plotted in Figure 5. The decrease of $H_C(T)$ reasonably follows a $\sqrt{T}$ law, as expected for uniaxial anisotropy $K$ within the framework of the coherent rotation model: these two hypotheses imply the variation of the energy barrier preventing magnetization reversal like $\Delta E \propto H^2$, and assuming a thermally activated magnetization switching with a relaxation rate obeying an Arrhenius law with a time constant $\tau = \tau_0 \exp(\Delta E/k_B T)$, one derives: $H_C = H_A \left[ 1 - \sqrt{\dfrac{\mathrm{Ln}(\tau/\tau_0) k_B T}{KV}} \right]$.

The blocking temperature determined by the intercept of this law with the $x$-axis is $T_B$=170 K, slightly lower than the value inferred from field-cooled / zero-field-cooled curves, see inset of Figure 5. Using the relationship $KV_n \approx 25 k_B T$, resulting from an attempt frequency $\tau_0 \approx 10^{-10}$ s and a measurement time $\tau \approx 10$ s, we infer $V_n \approx 100\,\mathrm{nm}^3$. Taking into account the finite thickness t=0.8nm, $V_n$ is readily converted into a surface around 130nm$^2$. If we assume that nucleation occurs in a single Fe wire at a time, of mean width 1.5nm, $V_n$ is in turn converted into a length of ≈90nm. This is significantly larger than the full wall width $\pi\sqrt{A/K} \leq 20$nm, which is an unusual situation. This large value suggests that nucleation events may imply two or more wires, via dipolar or RKKY coupling, or at the intersection of wires resulting from sample imperfections. This conclusion is supported by the rapid magnetization reversal evidenced at $H_C$ even at low temperature, which is usually the signature of a nucleation-propagation magnetization reversal process, incompatible with the picture of isolated wires reversing independently.

In conclusion, self-organized arrays of laterally-alternating (Fe,Ag) nanostripes with a period 3.5nm have been fabricated on Mo(110)/ $Al_2O_3(11\bar{2}0)$ using UHV-Pulsed Laser Deposition, up to a maximum thickness of 4 atomic layers. These arrays display a strong in-plane uniaxial anisotropy (K≈5.6±0.5×$10^5$J/$m^3$) along [001], the direction of the wires, resulting from a balance between dipolar, interface and magneto-elastic anisotropy. The wires are superparamagnetic at 300K, with a blocking temperature of 170-200K. Nucleation volumes are surprisingly large if analyzed in terms of wire length, suggesting the existence of interactions between the wires.

**Acknowledgments**

This work was partly supported by a Région-Rhône-Alpes grant. The authors wish to acknowledge V. Santonacci and Ph. David for technical advice and support.

**Figures**

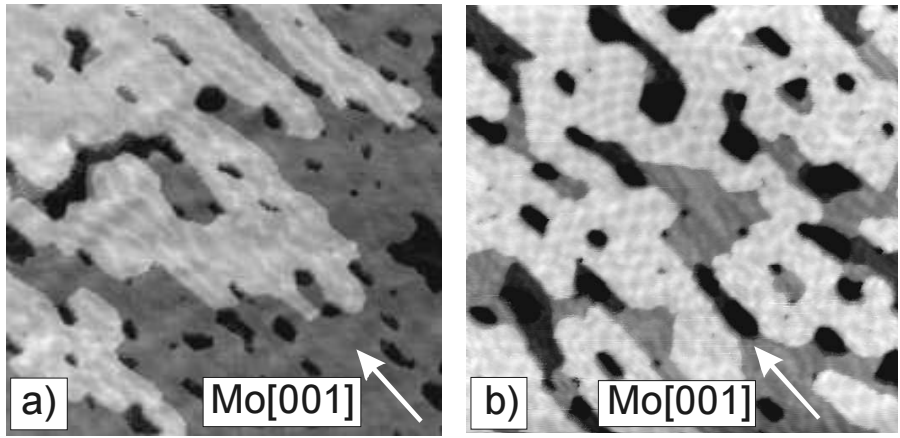

Fig.1. STM images (80 nm × 80 nm) showing the evolution of the morphology of the nanostructures obtained by sequential deposition of (0.5 AL Ag + 0.5 AL Fe)$_n$ on Mo(110) at 420K. **(a)** *n*=2, formation of stripes parallel to [001]Mo. **(b)** *n*=6, onset of a 2D relaxation network.

Fig. 2. Morphology of 4 ALs Fe/Ag nanostripes (200 nm x 200 nm STM image). The alternate Fe and Ag stripes are shown in a grey scale as dark and bright respectively. Topmost inset: thickness corrugation profile along the black line, giving the relative height of these stripes. Lower part inset: zoom of the area inside the circle, showing in detail the dark and bright grey corrugation.

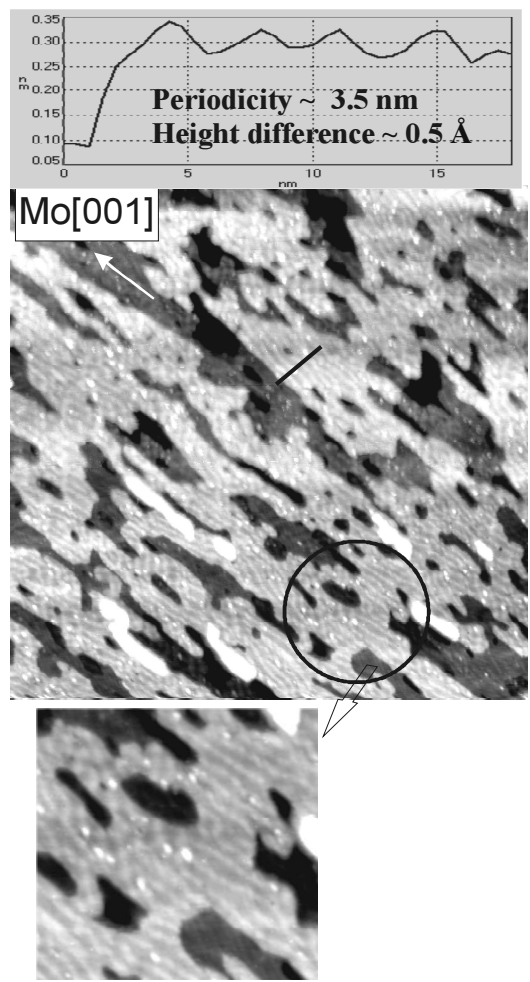

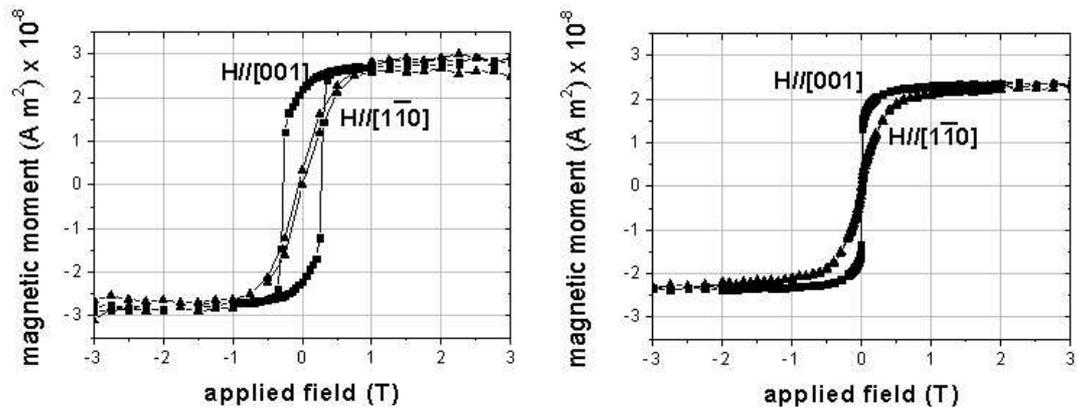

Fig.3. Hysteresis loops of (Fe,Ag)(4AL) nanostripes at 10K (left) and 300K (right) with an in-plane applied field parallel and perpendicular to [001] Mo (stripes' direction).

Fig. 4. Hysteresis loops of (Fe,Ag)(4AL) nanostripes as a function of temperature in an applied magnetic field parallel to Mo[001]

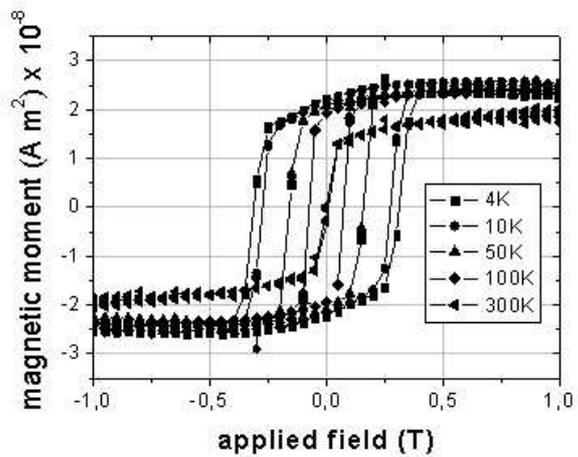

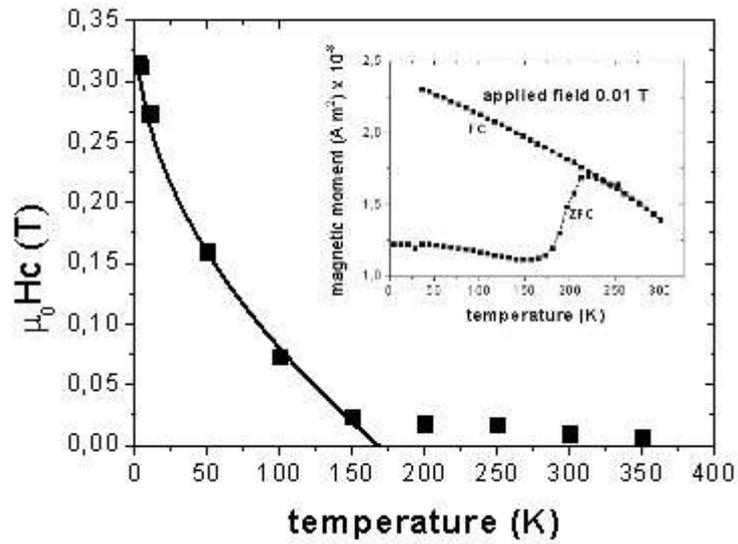

Fig. 5. Temperature dependence of the switching field $H_C$. The full curve is the result of the fit according to $T^{1/2}$. Inset: variation of the low-field magnetization, as a function of temperature, within a "zero-field cooling / Field-cooling" process. The applied field is 0.01T.